\documentclass{ws-procs11x85}
\usepackage{multicol}
\makeindex
\begin{document}

\title{LIKELIHOOD RATIO INTERVALS WITH BAYESIAN TREATMENT OF UNCERTAINTIES: COVERAGE, POWER AND COMBINED EXPERIMENTS}

\author{J. CONRAD}

\address{CERN\\ PH-EP Dept.\\ CH-1211 Geneva 23, Switzerland 
\\E-mail: Jan.Conrad@cern.ch}

\author{F. TEGENFELDT}

\address{Iowa State University \\ Ames, IA 5011-3160, USA \\E-mail: Fredrik.Tegenfeldt@cern.ch}

%%%%%%%%%%%%%%%%%%%%%%%%%%%%%%%%%%%%%%%%%%%%%%%%%%%%%%%%%%%%%%%%%%%%%%%%%
% You may repeat \author \address as often as necessary             %
%%%%%%%%%%%%%%%%%%%%%%%%%%%%%%%%%%%%%%%%%%%%%%%%%%%%%%%%%%%%%%%%%%%%%%%%%

\maketitle

\abstract{ In this note we present studies of coverage and power for confidence intervals for a Poisson process with known background calculated using the Likelihood ratio (aka Feldman \& Cousins) ordering with Bayesian treatment of uncertainties in nuisance parameters. We consider both the variant where the Bayesian integration is done in both the numerator and the denominator and the modification where the integration is done only in the numerator whereas in the denominator the likelihood is taken at the maximum likelihood estimate of the parameters.
Furthermore we discuss how measurements can be combined in this framework and give an illustration with limits on the branching ratio of a rare B-meson decay recently presented by CDF/D0. A set of C++ classes has been developed which can be used to calculate confidence intervals for single or combining multiple experiments using the above algorithms and considering a variety of parameterizations to describe the uncertainties.}

\begin{multicols}{2}

\baselineskip=13.07pt
\voffset-1.2in 
\section{Introduction}
A popular technique to calculate confidence intervals in recent years is the technique suggested by Feldman \& Cousins\cite{Feldman:1998a}. The method consists of constructing an acceptance region for each possible hypothesis (in the way as proposed by Neyman\cite{Neyman:1937a}) and fixing the limits of the region by including experimental outcomes according to rank which is given by the likelihood ratio\footnote{throughout this note we consider Poisson distributions with experimental outcome $n$, hypothesis parameter $s$ and (possibly not exactly) known background $b$}:
\begin{equation}
R(s,n)_{\mathcal{L}} = \frac{\mathcal{L}(n|s+b)}{\mathcal{L}(n|s_{best}+b)}
\end{equation}
where $s$ is the hypothesis, $n$ the experimental outcome, $b$ the expected background, $s_{best}$ is the  hypothesis most compatible with $n$ and $\mathcal{L}$ the Likelihood function. The expected background $b$ is an example for a so called {\it nuisance parameter.}, i.e. a parameter which is not of primary interest but which still affects the calculated confidence interval. Another example of such a nuisance parameter could be the signal efficiency. In the originally proposed method by Feldman \& Cousins, only the presence of background was considered and it was assumed to be exactly known. The question on how to treat uncertainties in nuisance parameters in confidence interval calculation, in particular in context of the frequentist construction has drawn considerable attention in the recent years. In 1992 Cousins \& Highland\cite{Cousins:1992a} proposed a method which is based on a Bayesian treatment of the nuisance parameters. The main idea is to use a probability density function (pdf) in which the average is taken over the nuisance parameter:
\begin{equation}
P(n|s,\epsilon) \longrightarrow \int P(n|s,\epsilon') P(\epsilon'|\epsilon)d\,\epsilon' := q(n|s,\epsilon)
\end{equation}
where $\epsilon'$ is the true value of the nuisance parameter, $\epsilon$ denotes its estimate and $s$ and $n$ symbolize the signal hypothesis and the experimental outcome respectively.

Cousins \& Highland only treated the case of Gaussian uncertainties in the signal efficiency. The method has since been generalized by Conrad et al.\cite{Conrad:2002kn} to operate with the Feldman \& Cousins ordering scheme and taking into account both efficiency and background uncertainties as well as correlations. This generalized method has already been  used in a number of particle and astroparticle physics experiments (see references in Tegenfeldt \& Conrad\cite{Tegenfeldt:2004dk}). FHC$^2$ denotes this generalized method in the remainder of this note.

In case of significantly less events observed than expected background, FHC$^2$ tends to result in  confidence intervals which are becoming smaller with increasing uncertainties. Hill\cite{Hill:2003jk} therefore proposed a modification where in the ordering the likelihood ratio is defined as:
\begin{equation}
R(s,n)_{\mathcal{L}} = \frac{q(n|s+b)}{\mathcal{L}(\max{(0,n_{obs}-\hat{b})}+\hat{b})}
\end{equation} 
here $\hat{b}$ is the maximum likelihood estimate of b given the subsidiary observation  of b. MBT (``Modified Bayesian Treatment'') denotes this modification in the remainder of this note.

In this contribution, we discuss coverage and power of these two methods as well as the combination of different experiments with and without correlations. We start by introducing the C++ library which has been developed to be able to do the necessary calculations.

\section{POLE++}
For the coverage studies presented in this paper a reasonably fast and efficient code is required. Hence, a user-friendly and flexible C++ library of classes was developed based on the FORTRAN routine presented by Conrad\cite{Conrad:Pole}. The library is independent of external libraries and consists of two main classes, {\em Pole} and {\em Coverage}. The first class takes as input the number of observed events, the efficiency and background with uncertainties and calculates the limits using the method described in this paper. The integrals are solved analytically. {\em Coverage} generates user-defined pseudo-experiments and calculates the coverage using {\em Pole}. Presently the library supports Gauss, log-Normal and flat pdf for description of the nuisance parameters. Several Experiments with correlated or uncorrelated uncertainties in the nuisance parameters can be combined. The pole++ library can be obtained from http://cern.ch/tegen/statistics.html

\section{Coverage and Power}
The most crucial property of methods for confidence interval construction is the coverage, which states that a fraction (1-$\alpha$) of  infinitely many repeated experiments should yield confidence intervals that include the true hypothesis irrespective of what the true hypothesis is.

For a confidence interval construction (according to Neyman) without uncertainties in nuisance parameters this property is fulfilled by construction. In the present case however, we have to test the coverage employing Monte Carlo experiments.

{\it Power} on the other hand is a concept which is defined in the context of hypothesis testing: the power of a hypothesis testing method is the probability that it will reject the null hypothesis, s$_0$, given that the alternative hypothesis s$_{true}$ is true. This concept is rather difficult to generalize to confidence intervals since the alternative hypothesis is not uniquely defined. We use the following definition for power:
\begin{equation}
\Pi(s_{true})_{s_0} = \sum_{n \notin Acc(s_0)}{q(n|s_{true},\epsilon)}
\end{equation}
and view power as a function of $s_{true}$. $Acc(s_0)$ here denotes the acceptance region of $s_0$. This seems an intuitively appealing measure: given the choice between different methods, the one should be taken which has minimally overlapping acceptance regions.

\begin{figure*}[htbp]
\center
\psfig{figure=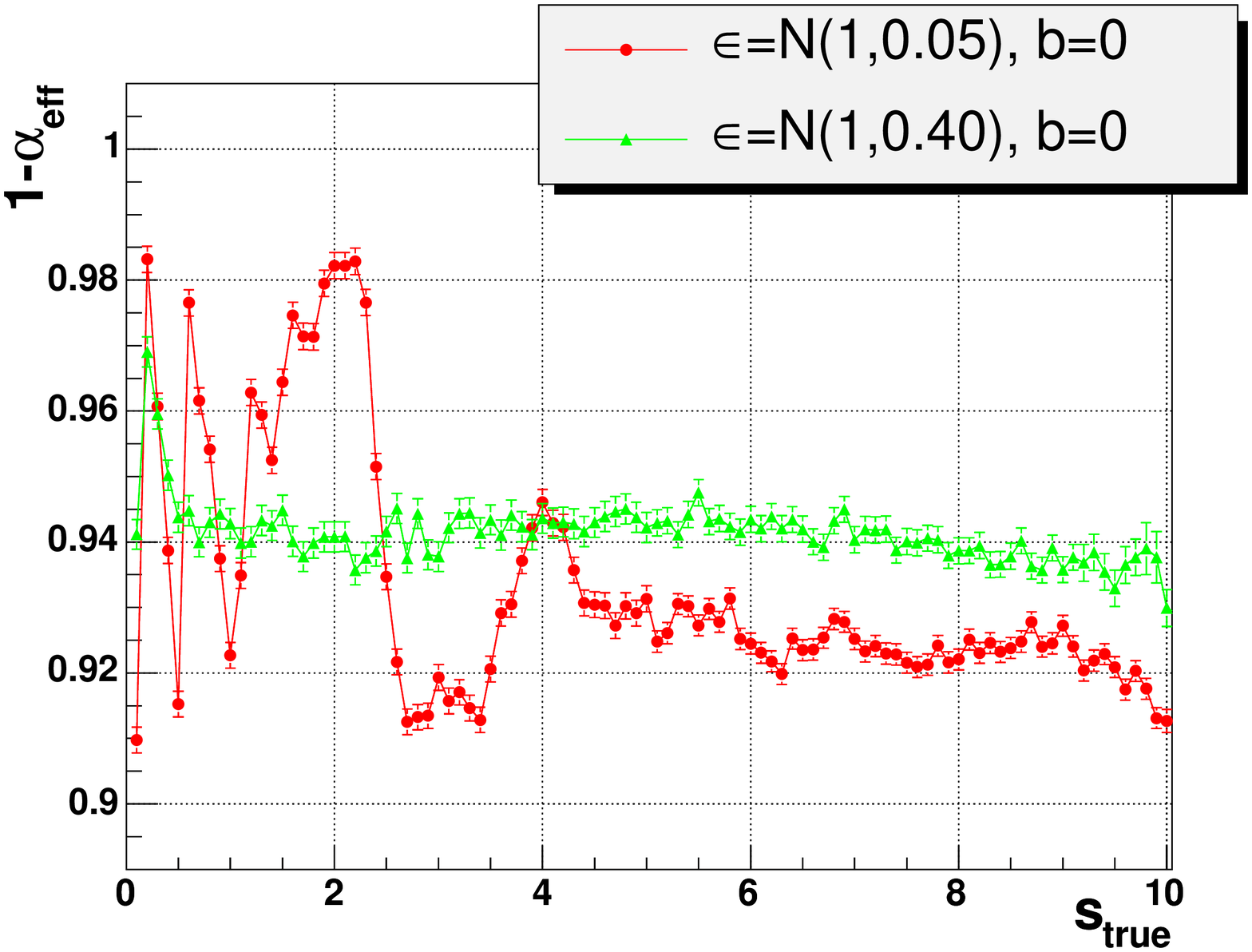,height=2.8in}
\psfig{figure=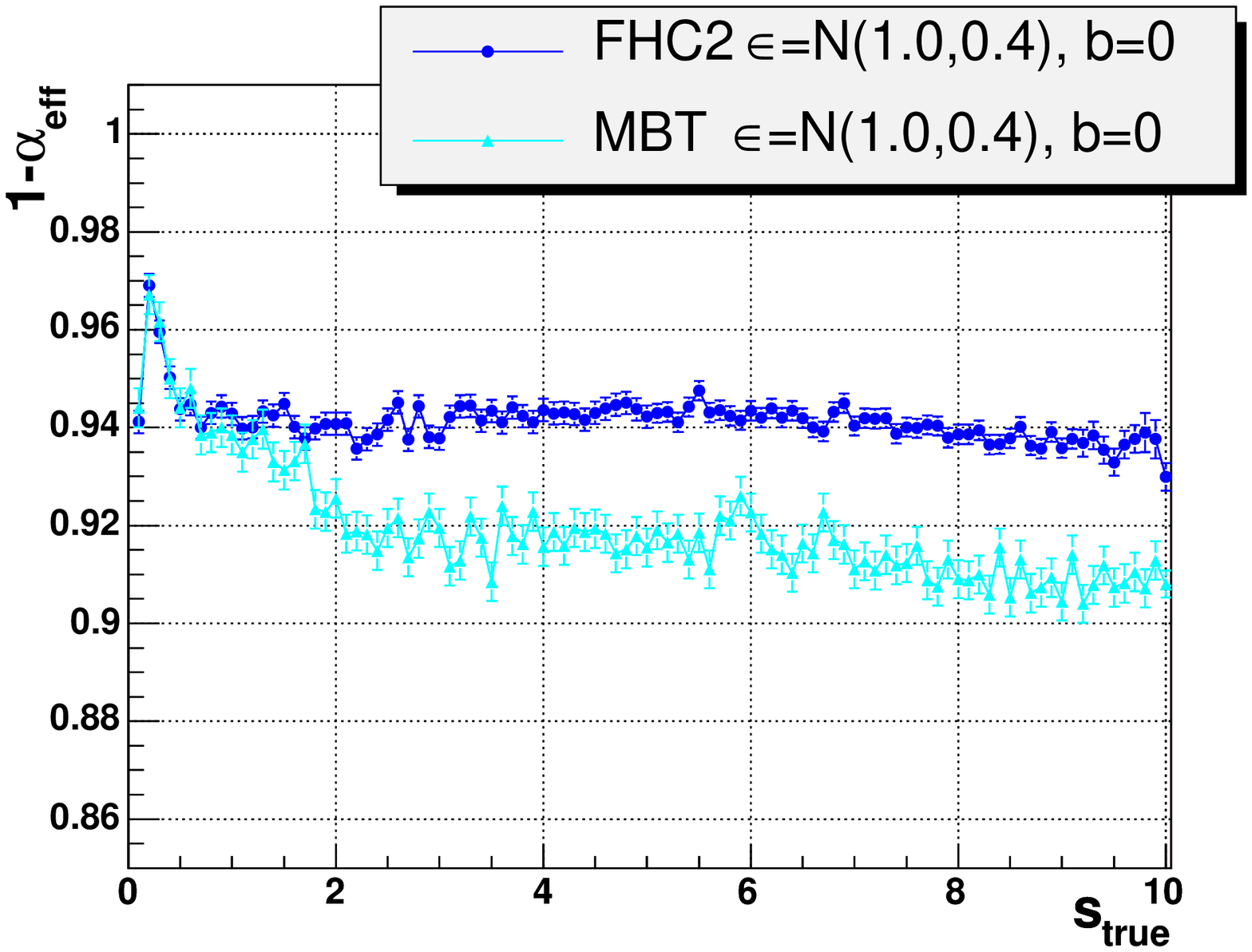,height=2.8in}
\psfig{figure=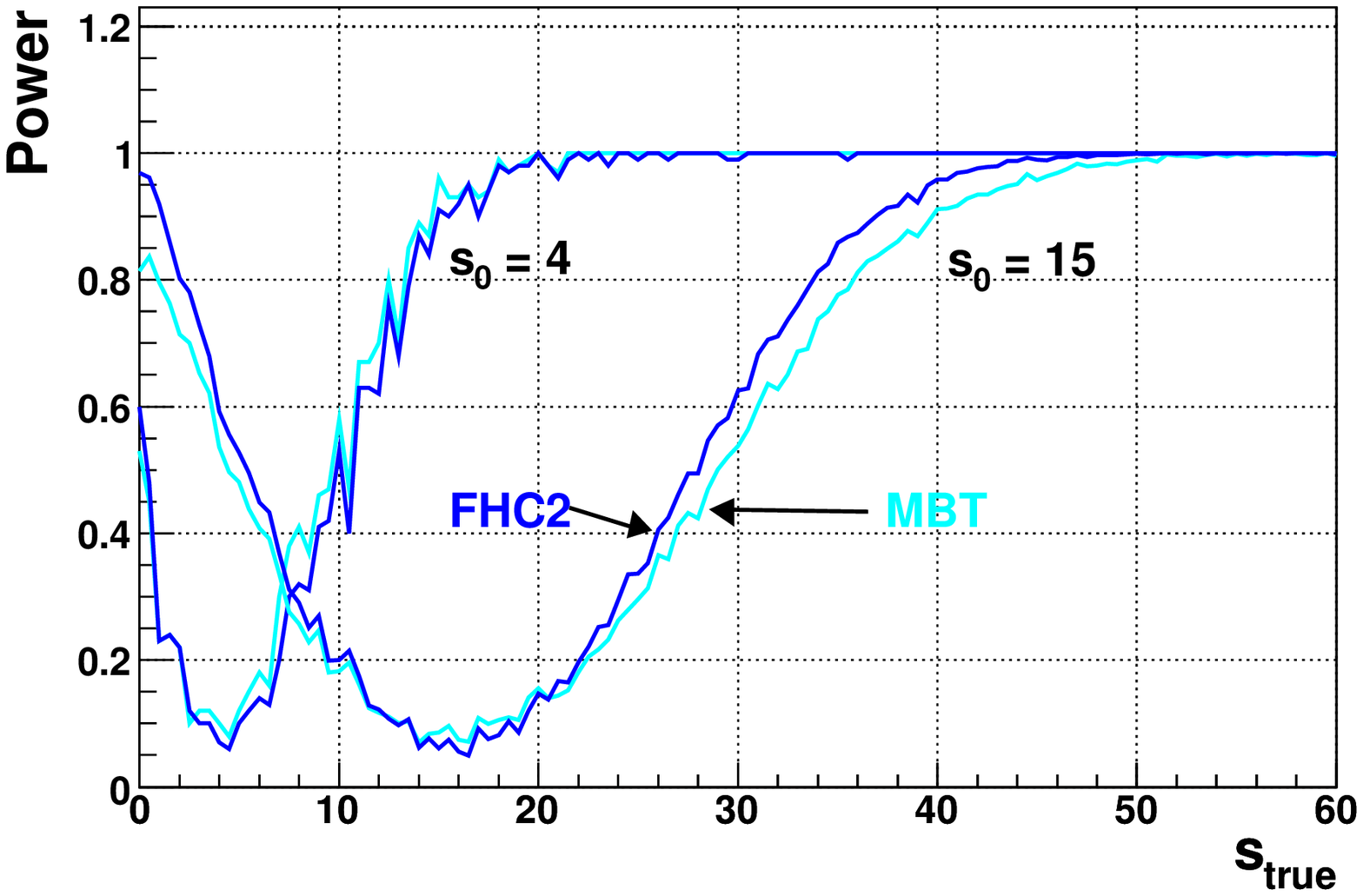,height=2.8in}
\caption{Examples for the coverage and power of the discussed methods. Upper most figure: coverage of the FHC$^2$ method assuming a 5 \% and 40 \% Gaussian uncertainties in efficiency. Middle figure: the coverage for the FHC$^2$ method compared to the MBT method for 40 \% Gaussian efficiency uncertainties. Lowest figure: the power of the two methods compared for 40 \% Gaussian uncertainties in efficiency.}
\label{fig:cov}
\end{figure*}

Typical examples of the coverage as function of signal hypothesis are shown in figure \ref{fig:cov}. It can be seen that the introduction of a continuous variable leads to a considerable smoothing of the coverage plot. A modest amount of over-coverage is introduced, similarly for the MBT method and the FHC$^2$ method. For high Gaussian uncertainties in efficiency ($\sim$ 40 \%) the over-coverage of MBT is less pronounced than for FHC$^2$. More detailed coverage studies  of the FHC$^2$ method have been presented by Tegenfeldt \& Conrad\cite{Tegenfeldt:2004dk}. The power of the  FHC$^2$ and MBT methods is compared in figure \ref{fig:cov} for 40 \% uncertainties in the efficiency. FHC$^2$ as higher power for hypotheses rather far away from the null hypotheses. This is true only for large signals and comparably large uncertainties (and for not too large differences between $s_0$ and $s_{true}$), otherwise differences are negligible.

\section{Combining different experiments}
The combination of experiments can be divided into two cases. The simpler case is the one of completely uncorrelated experiments: in this case the pdf used in the construction are given by a multiplication of the pdfs of the single experiments:
\begin{equation}
q(\vec{n}|s) = \prod_{i=1}^{n_{exp}} q(n_i|s,\epsilon_i)
\end{equation}
If correlations between uncertainties in nuisance parameters have to be considered, multivariate pdfs have to be employed:
\begin{equation}
q(\vec{n}|s,\vec{\epsilon}) = \int_0^\infty ...\int_0^\infty \prod_{i=1}^{n_{exp}} P(n|s,\epsilon'_i) P(\vec{\epsilon'}|\vec{\epsilon}) \prod_{i=1}^{n_{exp}} d\epsilon_i'
\end{equation}
We illustrate the effect of combining different experiments with the example of the CDF  limit on the branching ratio for $B^0_s \rightarrow \mu^+\mu^-$, see table \ref{tab:1}. In this case, two CDF data sets are combined with an uncorrelated uncertainty in the background expectation and an uncertainty in the efficiency which can be factorized into a correlated and uncorrelated part\cite{Bernhard:2005yn}. Bernhard et. al.\cite{Bernhard:2005yn} presented a fully Bayesian combination, which is included in the table for comparison. The limit obtained using the FHC$^2$ method is slightly smaller than the fully Bayesian limit.
\begin{tablehere}
\begin{center}
\caption{The CDF single and combined limits on $B^0_s \rightarrow \mu^+\mu^-$ calculated by FHC$^2$. CDF1 and CDF2 denote the two different data sets used for single limits. The quoted uncertainties are for the single experiments, the efficiency uncertainties change to 13.1 and 11.1 \% for the uncorrelated part if experiments are combined. The number in the parentheses is the result of the purely Bayesian calculation$^7$. \label{tab:1}} \vspace{0.2cm}
\begin{tabular}{|c|cc|c}
\hline
\raisebox{0pt}[12pt][6pt]{} &
\raisebox{0pt}[12pt][6pt]{CDF 1} &
\raisebox{0pt}[12pt][6pt]{CDF 2} \\
\hline
\raisebox{0pt}[12pt][6pt]{eff. uncertainty [\%]} &
\raisebox{0pt}[12pt][6pt]{18.2} &
\raisebox{0pt}[12pt][6pt]{16.0} \\
\hline
\raisebox{0pt}[12pt][6pt]{eff. uncertainty [\%]} &
\raisebox{0pt}[12pt][6pt]{20.3} &
\raisebox{0pt}[12pt][6pt]{19.2} \\
\hline
\raisebox{0pt}[12pt][6pt]{corr. eff. uncertainty.[\%] } &
\multicolumn{2}{|c|}{\raisebox{0pt}[12pt][6pt]{15.5} } \\
\hline

\raisebox{0pt}[12pt][6pt]{95 \% CL [10$^{-7}$]} &
\raisebox{0pt}[12pt][6pt]{2.5} &
\raisebox{0pt}[12pt][6pt]{4.3}\\ 
\hline
%\cline{1-2}
\raisebox{0pt}[12pt][6pt]{95 \% comb.[10$^{-7}$] } &
\multicolumn{2}{|c|}{\raisebox{0pt}[12pt][6pt]{1.7 (2.0)} } \\
\hline
%\cline{1-3}
\end{tabular}
\end{center}
\end{tablehere}

\section{Discussion \& Conclusion}
There are two main caveats when interpreting the presented results: first of all, the methods (more or less implicitly) assume a flat prior probability for the true nuisance parameter. Thus, conclusions on the coverage and power are true only for that prior. This assumption seems particularly harmful in case of combined experiments, a case for which we did not calculate the coverage. Results presented at this conference by Heinrich\cite{Heinrich:2005} indicate that the assumption of a flat prior for nuisance parameters in each channel leads to significant under-coverage for fully Bayesian confidence intervals. Heinrich also shows, that this behavior can be remedied with an appropriate choice of prior (in his particular example: 1/$\epsilon$). For the methods presented here this might imply that there is under-coverage in case of several combined experiments. A second caveat, is that we test the coverage only for 90\% confidence level. At this conference Cranmer\cite{Cranmer:2005} presented results that indicate under-coverage for very high confidence levels ($>$ 5 $\sigma$) if uncertainties in the background are treated in the Bayesian way. Tests of coverage for high confidence levels and combined experiments are currently under way. With these caveats in mind, we conclude that Bayesian treatment of nuisance parameters introduces a moderate amount of over-coverage. The MBT method has less over-coverage for the case with large Gaussian uncertainties in the signal efficiencies. We also compared the power of the two suggested methods. For large uncertainties and large true signals, the FHC$^2$ method has higher power for hypotheses relatively far away from the null hypothesis. 

\baselineskip=13.07pt

\section*{Acknowledgments}
We would like to thank the conference organizers, in particular Louis Lyons for organizing this useful and very enjoyable conference.

\end{multicols}

\begin{thebibliography}{99}
\raggedright
\bibitem{Feldman:1998a} G.~J.~Feldman and R.~D.~Cousins, Phys. Rev {\bfseries D57},  3873, (1998).
\bibitem{Cousins:1992a} R.~D.~Cousins and V.~ L.~Highland, Nucl. \ Instrum. \ Meth. A {\bf 320} (1992) 331.
\bibitem{Neyman:1937a} J.~Neyman, Phil. Trans. Royal Soc. London {\bfseries A}, 333, (1937).
\bibitem{Conrad:2002kn}
J.~Conrad, O.~Botner, A.~Hallgren and C.~Perez de los Heros,
%``Including systematic uncertainties in confidence interval construction  for
%Poisson statistics,''
Phys.\ Rev.\ D {\bf 67} (2003) 012002
%[arXiv:hep-ex/0202013].
%%CITATION = HEP-EX 0202013;%%
%\cite{Tegenfeldt:2004dk}
\bibitem{Tegenfeldt:2004dk}
F.~Tegenfeldt and J.~Conrad,
%``On Bayesian treatment of systematic uncertainties in confidence  interval
%calculations,''
Nucl.\ Instrum.\ Meth.\ A {\bf 539} (2005) 407
[arXiv:physics/0408039].
%%CITATION = PHYS-ICS 0408039;%%
%\cite{Hill:2003jk}
\bibitem{Hill:2003jk}
G.~C.~Hill,
%``Comment on 'Including systematic uncertainties in confidence interval
%construction for Poisson statistics',''
Phys.\ Rev.\ D {\bf 67}, 118101 (2003)
%[arXiv:physics/0302057].
%%CITATION = PHYS-ICS 0302057;%%
%\bibitem{Conrad:2002ur} J.~Conrad, O.~Botner, A.~Hallgren and C.~Perez de los Heros, published in Proc. of Conference on Advanced Statitical Techniques in Particle Physics, Durham, England, March 2002

%\cite{Bernhard:2005yn}
\bibitem{Bernhard:2005yn}
R.~Bernhard {\it et al.}  [CDF Collaboration],
%``A Combination of {CDF} and {D0} Limits on the Branching Ratio of $B^0_s(d)
%\to \mu^+ \mu^-$ Decays,''
arXiv:hep-ex/0508058.
%%CITATION = HEP-EX 0508058;%%
\bibitem{Conrad:Pole}  J.~Conrad, Comp. Phys. Comm. {\bfseries 158} 117 (2004)
\bibitem{Heinrich:2005}  J.~Heinrich, these proceedings
\bibitem{Cranmer:2005}  K.~Cranmer, these proceedings

\end{thebibliography}
\end{document}